\documentclass[conference]{IEEEtran}
\IEEEoverridecommandlockouts

\usepackage{cite}
\usepackage{amsmath,amssymb,amsfonts}
\usepackage{algorithmic}
\usepackage{graphicx}
\usepackage{textcomp}
\usepackage{xcolor}
\usepackage{booktabs}
\usepackage{multirow}
\usepackage{hyperref}
\usepackage{tabularx}
\usepackage{eso-pic}   

\def\BibTeX{{\rm B\kern-.05em{\sc i\kern-.025em b}\kern-.08em
    T\kern-.1667em\lower.7ex\hbox{E}\kern-.125emX}}

\begin{document}


\title{MILD: \textbf{M}ulti-\textbf{I}ntent \textbf{L}earning and \textbf{D}isambiguation for Proactive Failure Prediction in Intent-based Networking}

\author{
\IEEEauthorblockN{
Md. Kamrul Hossain\IEEEauthorrefmark{1},
Walid Aljoby\IEEEauthorrefmark{2}
}
\IEEEauthorblockA{
\IEEEauthorrefmark{1}\IEEEauthorrefmark{2}Information and Computer Science Department, King Fahd University of Petroleum and Minerals, Dhahran 31261, Saudi Arabia\\
\IEEEauthorrefmark{2}IRC for Intelligent Secure Systems, King Fahd University of Petroleum and Minerals, Dhahran 31261, Saudi Arabia
}
\IEEEauthorblockA{(e-mail: g202215400@kfupm.edu.sa, waleed.gobi@kfupm.edu.sa)
}
}


\maketitle

\AddToShipoutPictureFG*{%
  \AtPageLowerLeft{%
    \raisebox{40pt}{%
      \makebox[\paperwidth]{%
        \centering\footnotesize
        © 2026 IEEE. Accepted for publication in IEEE/IFIP NOMS 2026.
      }%
    }%
  }%
}

\begin{abstract}
In multi-intent intent-based networks, a single fault can trigger co-drift where multiple intents exhibit symptomatic KPI degradation, creating ambiguity about the true root-cause intent. We present MILD, a proactive framework that reformulates intent assurance from reactive drift detection to fixed-horizon failure prediction with intent-level disambiguation. MILD uses a teacher-augmented Mixture-of-Experts where a gated disambiguation module identifies the root-cause intent while per-intent heads output calibrated risk scores. 
On a benchmark with non-linear failures and co-drifts, MILD provides 3.8\%--92.5\% longer remediation lead time and improves intent-level root-cause disambiguation accuracy by 9.4\%--45.8\% over baselines.
MILD also provides per-alert KPI explanations, enabling actionable diagnosis.

\end{abstract}


\section{Introduction}
\label{sec:intro}
Cloud-native 5G/6G and modern enterprise/data-center networks increasingly require operating heterogeneous services under strict SLAs, making device-level, imperative configuration brittle and hard to scale \cite{10529727}. This motivates \textbf{Intent-Based Networking (IBN)} \cite{9925251,rfc9315}, where operators specify \emph{what} the network should achieve using high-level intents that are translated into low-level configurations and continuously verified via telemetry-driven closed-loop assurance.

A critical component of IBN is \textbf{intent assurance} \cite{10575429} which is continuously checking whether observed KPIs comply with the intended goals. Intent failures are often preceded by subtle \textbf{intent drift} \cite{rfc9315}, i.e., gradual KPI deviations that are hard to detect early. Existing approaches raise alarm only when degradation is already severe. More importantly, they struggle in \textbf{multi-intent} settings where intents interact and a single fault can induce \emph{co-drift} across multiple intents, yielding ambiguous KPI symptoms that obscure the true \emph{root-cause} intent and hinder proactive mitigation \cite{11073595,10575429,10001426}. To illustrate, consider co-located microservices that implement multiple intents (e.g., API gateway, ingestion pipeline, analytics) while sharing CPU and memory \cite{10.1145/3501297,10.1145/3580305.3599934}. A fault such as CPU contention can degrade one service (root cause) and trigger cascading KPI symptoms in the others (victims), producing co-drifting KPI patterns that confound conventional drift detectors.

We address this by reformulating intent assurance from reactive drift detection to proactive \textbf{fixed-horizon failure prediction} with \textbf{intent-level disambiguation}. We propose \textbf{MILD} (Multi-Intent Learning and Disambiguation), a teacher-augmented Mixture-of-Experts (MoE) architecture \cite{chen2022towards} trained with a hybrid objective that jointly (i) predicts whether an intent will violate its goal within horizon $H$ and (ii) identifies the root-cause intent under co-drift. Beyond early warning, MILD provides operational insight via per-alert \textbf{KPI attribution} (via SHAP\cite{10.5555/3295222.3295230}) and supports multi-horizon deployment to bound time-to-failure. In Sec.~\ref{sec:exp}, we show evaluation of MILD on a reproducible benchmark with explicitly modeled co-drift scenarios. Results show that MILD achieves perfect detection while delivering earlier warning (lead time), lower false positives, and more accurate intent-level root-cause disambiguation than competitive supervised and drift-score baselines.


\section{Related Work}
Recent work such as NetIntent leverages LLMs to assist intent realization \cite{11293797}, but its reliance on active tests (e.g., iperf, ping) can be hard to scale and does not directly resolve root-cause ambiguity under co-drift. A related line of work forecasts individual KPIs or near-term network states using sequence models or classifiers (e.g., LSTM-based forecasting \cite{10456766,8891022,9615580} and SVM-based congestion prediction \cite{xing2023overview,10942926}), but these approaches lack an explicit mechanism to disambiguate which intent is the root cause when multiple intents exhibit correlated symptoms \cite{10.1145/3580305.3599934,Zanouda_2024}. Other works formalize intent drift via unsupervised anomaly detection on telemetry (e.g., DBSCAN clustering \cite{10770652}) or define drift as deviation from a target KPI state and use LLMs to propose corrective actions \cite{10575429,11073595}; while effective for detection/repair, these methods are largely reactive and do not explicitly learn fixed-window precursors or provide time-to-failure bounds. Furthermore, in contrast to NWDAF-style analytics functions that primarily support monitoring/diagnosis workflows\cite{9824403,ardestani2025nwdafenabledanalyticsclosedloopautomation}, MILD targets fixed-horizon proactive prediction with explicit multi-intent root-cause disambiguation under co-drift. Fixed-horizon prediction has been explored for general failures \cite{Basikolo2023TowardsZD,9557387}, but its use for multi-intent IBN assurance in jointly predicting failures and disambiguating root-cause intent remains underexplored. Our work fills this gap by reformulating assurance as fixed-horizon multi-intent failure prediction with intent-level disambiguation, operationalized via a teacher-augmented MoE trained to separate root-cause from victim intents under co-drift.

\section{The MILD Framework}
\label{sec:framework}
\begin{figure*}[t!]
    \centering
    \includegraphics[width=\linewidth]{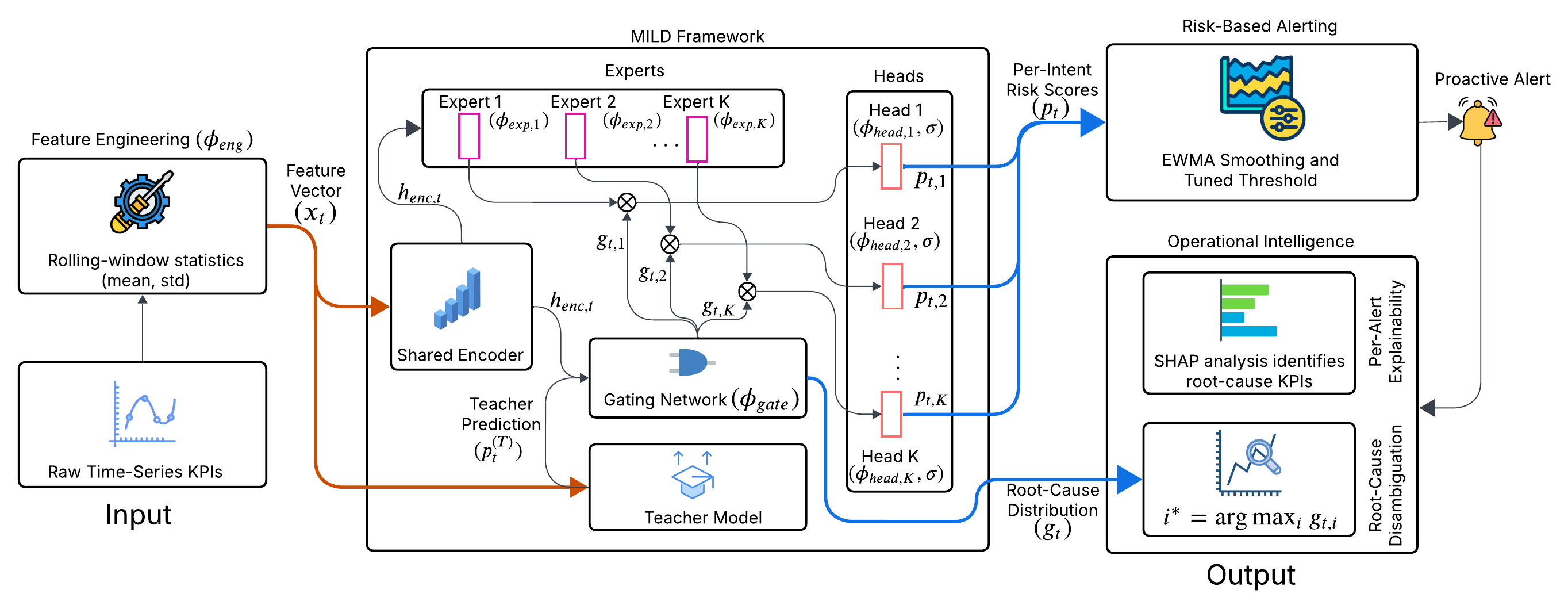}
    \caption{MILD: teacher-augmented MoE with risk prediction, root-cause disambiguation, and per-alert intelligence.}
    \label{fig:mild_framework}
\end{figure*}

MILD addresses \emph{multi-intent} proactive assurance where a single fault can induce \emph{co-drift} across intents, making the root cause ambiguous. We cast assurance as supervised learning that jointly outputs (i) per-intent fixed-horizon failure risks and (ii) a root-cause distribution over intents. MILD implements this via a teacher-augmented MoE trained with a hybrid objective where at runtime, risks are smoothed into alerts and explained per alert. Fig.~\ref{fig:mild_framework} shows MILD’s flow from KPI features and teacher priors to per-intent risks, root-cause gating, and EWMA-threshold alerting with SHAP explanations.

\subsection{Problem Formulation and Fixed-Horizon Labels}
\label{subsec:labeling}
We monitor $K$ intents $\mathcal{I}=\{1,\ldots,K\}$ using a KPI stream $\{\mathbf{k}_t\}$. At each time $t$, engineered features $\mathbf{x}_t=\phi_{\mathrm{eng}}(\mathbf{k}_{t-\Delta:t})\in\mathbb{R}^{D'}$ are computed from a recent window. For each intent $i$, let $t_{\mathrm{fail},i}$ be the (annotated) failure time. Given horizon $H$ (minutes), MILD outputs a risk score $p_{t,i}\in[0,1]$ and a root-cause distribution $\mathbf{g}_t$ with $\sum_i g_{t,i}=1$.

Training uses fixed-horizon labels:
\begin{equation}
\label{eq:bin_label}
y^{\mathrm{bin}}_{t,i}=\mathbf{1}\{t_{\mathrm{fail},i}-H \le t < t_{\mathrm{fail},i}\},
\end{equation}
and a root-cause indicator within the positive window,
\begin{equation}
\label{eq:cause_label}
y^{\mathrm{cause}}_{t,i}=\mathbf{1}\{\text{intent } i \text{ is the annotated root cause at } t\}\cdot y^{\mathrm{bin}}_{t,i}.
\end{equation}
When root-cause supervision is missing at $t$ (all-zero $\mathbf{y}^{\mathrm{cause}}_t$), we use $\mathbf{y}^{\mathrm{bin}}_t$ as weak gate supervision; for Kullback-Leibler (KL) divergence-based training, the supervision vector is normalized to sum to one.

\subsection{Teacher-Augmented MoE and Hybrid Training}
\vspace{-0.3cm}
\begin{equation}
\label{eq:mapping}
f_\theta:(\mathbf{x}_t,\mathbf{d}^{(T)}_t)\mapsto(\mathbf{p}_t,\mathbf{g}_t),
\end{equation}
where $\mathbf{p}_t=\{p_{t,i}\}$ and $\mathbf{d}^{(T)}_t$ is a teacher-derived distribution over intents. The teacher is a One-vs-Rest (OvR) Logistic Regression trained per intent on standardized features to estimate $p^{(T)}_{t,i}=\Pr(y^{\mathrm{bin}}_{t,i}=1\mid \mathbf{x}_t)$; we form $\mathbf{d}^{(T)}_t$ by applying temperature scaling to the teacher logits and a softmax over intents.

\textbf{Model:} A shared encoder produces $\mathbf{h}_{\mathrm{enc},t}=\phi_{\mathrm{enc}}(\mathbf{x}_t)$. The gate outputs:
\begin{equation}
\label{eq:gate}
\mathbf{g}_t=\mathrm{softmax}\!\left(\phi_{\mathrm{gate}}\left([\mathbf{h}_{\mathrm{enc},t};\mathbf{d}^{(T)}_t]\right)\right).
\end{equation}
Each expert--head branch yields the per-intent risk
$p_{t,i}=\sigma(\phi_{\mathrm{head},i}(g_{t,i}\cdot \phi_{\mathrm{exp},i}(\mathbf{h}_{\mathrm{enc},t})))$,
where $\sigma(\cdot)$ is the logistic sigmoid.

\textbf{Loss:} Each head minimizes a weighted combination of a focal-style classification loss \cite{lin2017focal} for $y^{\mathrm{bin}}_{t,i}$ and a distillation term \cite{hinton2015distilling} that matches the teacher probability $p^{(T)}_{t,i}$:
\begin{equation}
\label{eq:headloss}
\mathcal{L}_{\mathrm{head},i}=\alpha\,\mathcal{L}_{\mathrm{focal},i} + (1-\alpha)\,\mathcal{L}_{\mathrm{distill},i}.
\end{equation}
The gate is trained for disambiguation via supervision, teacher guidance, and sparsity:
\begin{equation}
\label{eq:gateloss}
\mathcal{L}_{\mathrm{gate}}=
w_c\,\mathrm{KL}( \bar{\mathbf{y}}_t \parallel \mathbf{g}_t)
+ w_T\,\mathrm{KL}( \mathbf{d}^{(T)}_t \parallel \mathbf{g}_t)
+ \lambda_s\sum_i g_{t,i}(1-g_{t,i}),
\end{equation}
where $\bar{\mathbf{y}}_t$ is the normalized gate-supervision vector (cause when available, otherwise bin). Optionally, we add a decorrelation regularizer over head representations \cite{cogswell2015reducing}. The total objective is:
\begin{equation}
\label{eq:totalloss}
\mathcal{L}_{\mathrm{total}}=\sum_{i=1}^K \mathcal{L}_{\mathrm{head},i} + \lambda_{\mathrm{gate}}\,\mathcal{L}_{\mathrm{gate}} \;(+\;\lambda_{\mathrm{decorr}}\,\mathcal{L}_{\mathrm{decorr}}).
\end{equation}

\subsection{Runtime Alerting and Operational Intelligence}
At inference, risks are smoothed with exponentially weighted moving average (EWMA)~\cite{Hyndman_Athanasopoulos_2021}, i.e., $\tilde{p}_{t,i}=\alpha_i p_{t,i}+(1-\alpha_i)\tilde{p}_{t-1,i}$ with $\alpha_i=\frac{2}{W_i+1}$. An alert fires when $\tilde{p}_{t,i}\ge \tau_i$. Here $W_i$ is the EWMA span and $\tau_i$ is the alert threshold (tuned on validation under a false-positive budget).

The root-cause intent at alert time $t$ is $i^{*}=\arg\max_i g_{t,i}$. We assume a single dominant root-cause intent per episode; extending $\mathbf{g}_t$ to multi-label root causes is a natural direction (e.g., by training the gate with multi-hot causes when available). For operator insight, we explain alerts using SHAP \cite{10.5555/3295222.3295230} on the alerted intent’s risk output. For urgency, we optionally deploy independent models with multiple horizons $\{H_1,\ldots,H_N\}$ to bound time-to-failure (TTF) (e.g., alert at $H_{\mathrm{long}}$ but not $H_{\mathrm{short}}$ implies $\widehat{\mathrm{TTF}}\in(H_{\mathrm{short}},H_{\mathrm{long}}]$.

\section{Experimental Results}
\label{sec:exp}
We evaluate MILD on the cloud-native use case mentioned in Sec.~\ref{sec:intro} with three intents (API, Telemetry, Analytics). Intents are monitored using the base KPIs in Table~\ref{tab:kpis}.

\begin{table}[t!]
\centering
\scriptsize
\caption{Base KPIs Selected for the Experimental Use Case}
\label{tab:kpis}
\begin{tabular}{@{}lp{3cm}l@{}}
\toprule
\textbf{KPI Name} & \textbf{Description} & \textbf{Category} \\ \midrule
\textit{cpu \%} & CPU utilization percentage & System Resource \\
\textit{ram \%} & RAM utilization percentage & System Resource \\
\textit{storage \%} & Storage utilization percentage & System Resource \\ \addlinespace
\textit{snet} & Service network throughput & Network Health \\
\textit{sri} & Service Reliability Index (a measure of availability) & Network Health \\ \addlinespace
\textit{api\_latency} & End-to-end latency for API requests & Application-Specific \\
\textit{telemetry\_queue} & Length of the data ingestion queue & Application-Specific \\
\textit{analytics\_tput} & Throughput of the data processing service & Application-Specific \\ \bottomrule
\end{tabular}
\end{table}

\subsection{Dataset Generation}
We generate a synthetic benchmark with 1-minute sampling that mimics realistic KPI behavior (seasonality, Gaussian noise) and includes unlabeled benign hard negatives (failure-like transients and high-load CPU/RAM spikes) to stress false positives \cite{10.1145/3501297}. We inject labeled failures in three categories \cite{10.1145/3501297, 10.1145/3580305.3599934}: (i) simple independent drifts, (ii) non-linear single-intent failures (multi-KPI interactions), and (iii) multi-intent co-drifts where one intent is labeled as root cause and another as symptomatic victim. The dataset contains 200{,}000 minute-level samples; events are weighted toward complex cases (60\% non-linear, 20\% co-drift, 20\% simple). We augment base KPIs with rolling-window statistics (5/15-min mean/std) and generate labels using the fixed-horizon strategy discussed in Sec.~\ref{subsec:labeling}.

\textbf{Prototype real-telemetry collection (PoC):} To validate feasibility beyond synthetic traces, we also built a small hardware-in-the-loop prototype that captures telemetry from (i) a P4 SmartNIC pipeline integrated with a Mininet/BMv2 testbed, and (ii) an HTTP microservice deployment where controlled load/drift episodes can be triggered. This prototype is not yet a complete, fully-labeled benchmark (hence not used as the primary evaluation dataset), but it demonstrates that MILD’s telemetry/feature pipeline can be driven by real measurements. The prototype scripts and artifacts are released alongside our code repository\cite{HossainMILD}.

\subsection{Implementation and Training Details}
\label{subsec:MILD_setting}
We use 10-fold blocked cross-validation (train on history preceding each test block) with an 80/20 train/validation split and early stopping (patience 8). MILD is trained with Adam (learning rate $10^{-3}$), batch size 512, up to 30 epochs. Key hyperparameters are $\alpha{=}0.9$, $T{=}2.0$, $w_c{=}0.7$, $w_T{=}0.7$, $\lambda_s{=}0.005$, and a false-positive budget of 1/day. All randomness is fixed with seed 42. Code and data are available on GitHub \cite{HossainMILD}.

\textbf{Overhead:} MILD’s inference cost scales linearly with the number of intents ($O(K)$ heads plus a shared encoder/gate). On our setup, we observe negligible batch inference overhead ($\approx$0.04 ms per sample) and $\approx$45 ms per minute-step in a real-time loop.

\subsection{Performance Analysis of MILD}
\label{subsec:results}

\subsubsection{Cross-validation outcome}
Table~\ref{tab:mild_performance} summarizes 10-fold results. MILD achieves 100\% detection for all intents with large lead times ($\approx$89--111 min). It maintains 3.90$\pm$3.74 false positives/day and achieves 88.67$\pm$7.96\% root-cause accuracy over all failure types (including co-drifts).

\begin{table}[!t]
\centering
\scriptsize
\caption{Performance of MILD from 10-Fold Blocked Cross-Validation}
\label{tab:mild_performance}
\begin{tabular}{@{}llc@{}}
\toprule
\textbf{Metric} & \textbf{Details} & \textbf{Value (mean $\pm$ std)} \\ \midrule
\multirow{3}{*}{Failure Detection Rate (\%)} & Analytics & $100.00 \pm 0.00$ \\
 & API & $100.00 \pm 0.00$ \\
 & Telemetry & $100.00 \pm 0.00$ \\ \cmidrule(l){2-3}
\multirow{3}{*}{Avg. Lead Time (min)} & Analytics & $88.79 \pm 8.32$ \\
 & API & $95.48 \pm 10.00$ \\
 & Telemetry & $110.93 \pm 11.02$ \\ \midrule
FP Rate per Day & Overall & $3.90 \pm 3.74$ \\ \midrule
Disambiguation Accuracy (\%) & Root Cause Accuracy & $88.67 \pm 7.96$ \\ \bottomrule
\end{tabular}
\end{table}

\textbf{Error analysis:} Most false positives arise from benign mimics (transient KPI excursions) that resemble partial drifts but self-resolve; most disambiguation errors occur early in co-drift episodes when the victim intent’s risk rises first before the gate consolidates on the true cause.

\subsubsection{Disambiguation}
\begin{figure}[t!]
    \centering
    \includegraphics[width=0.75\columnwidth]{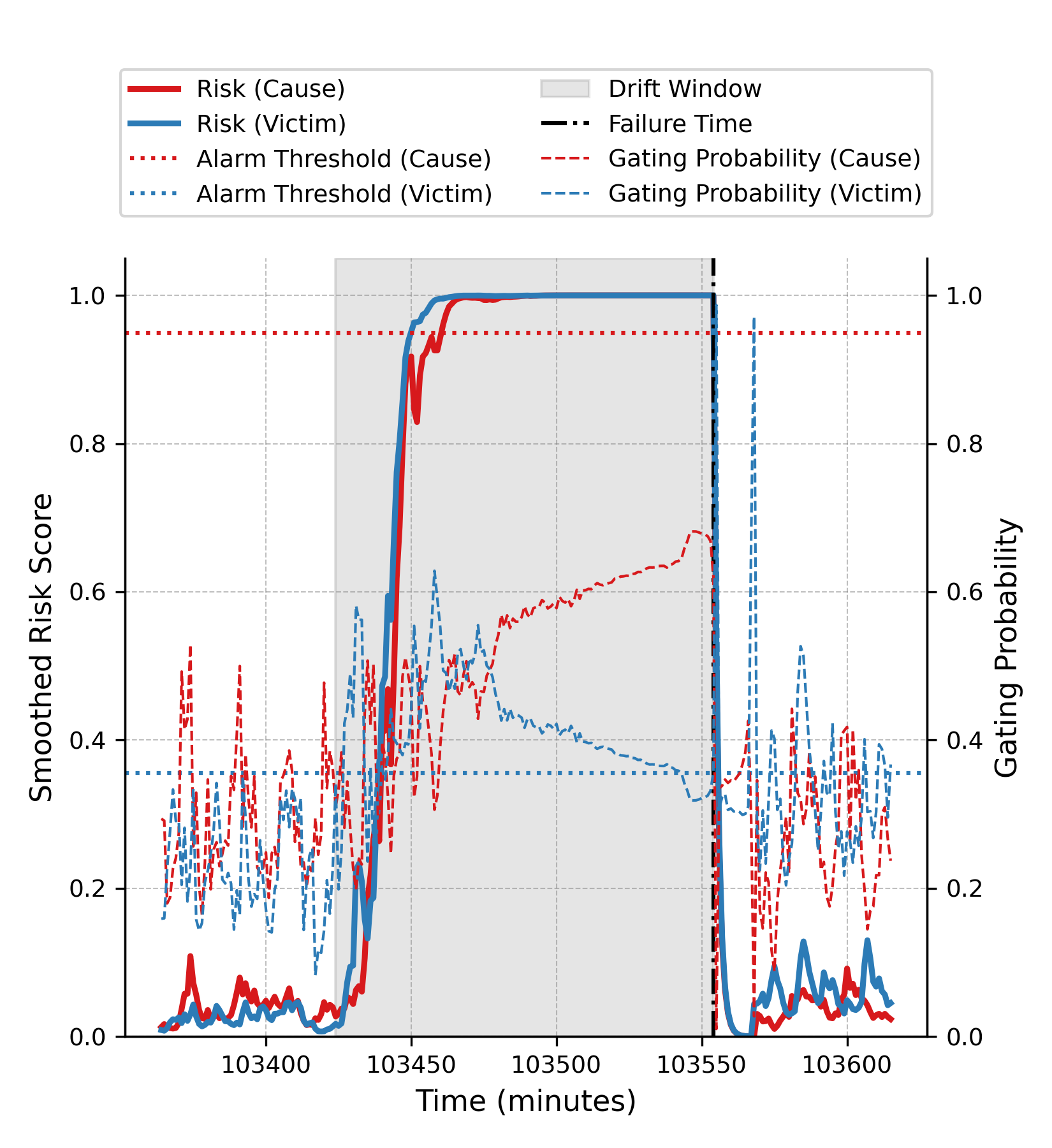}
    \caption{MILD's disambiguation during a co-drift event. Smoothed risk scores (left y-axis) and gating probabilities (right y-axis) are shown for the root cause (analytics, red) and the victim (telemetry, blue) intents.}
    \label{fig:disambiguation_example}
\end{figure}

Fig.~\ref{fig:disambiguation_example} shows a representative co-drift where an \textit{analytics} fault (root cause) induces \textit{telemetry} symptoms. Although the victim risk rises early, the gate probability shifts toward the true cause well before failure, illustrating robust root-cause disambiguation under evolving signals.

\subsubsection{Per-Alert Explainability via SHAP}
\begin{figure}[t!]
    \centering
    \includegraphics[width=0.75\columnwidth]{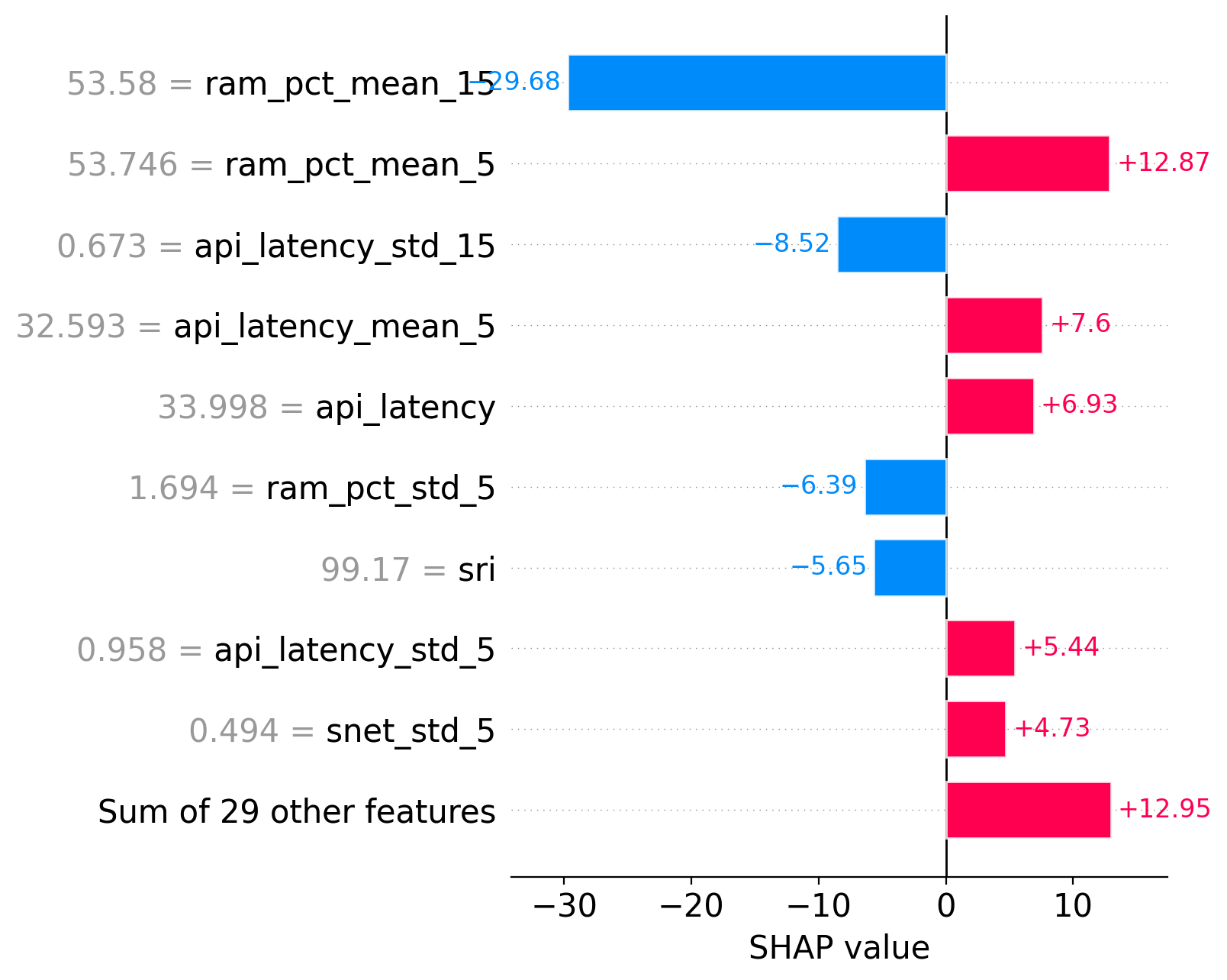}
    \caption{Example SHAP explanation for a telemetry alert (feature contributions to risk score).}
    \label{fig:shap}
\end{figure}

Each alert is explained using SHAP to highlight the features most responsible for the predicted risk. Fig.~\ref{fig:shap} shows an example alert on \textit{telemetry} (128 min before failure), where recent RAM and API-latency features dominate the positive contribution despite being moderated by long-term RAM stability, providing actionable KPI-level guidance.

\subsubsection{Dynamic Lead Time Estimation with Multi-Horizon Models}
\begin{figure}[t!]
    \centering
    \includegraphics[width=0.9\columnwidth]{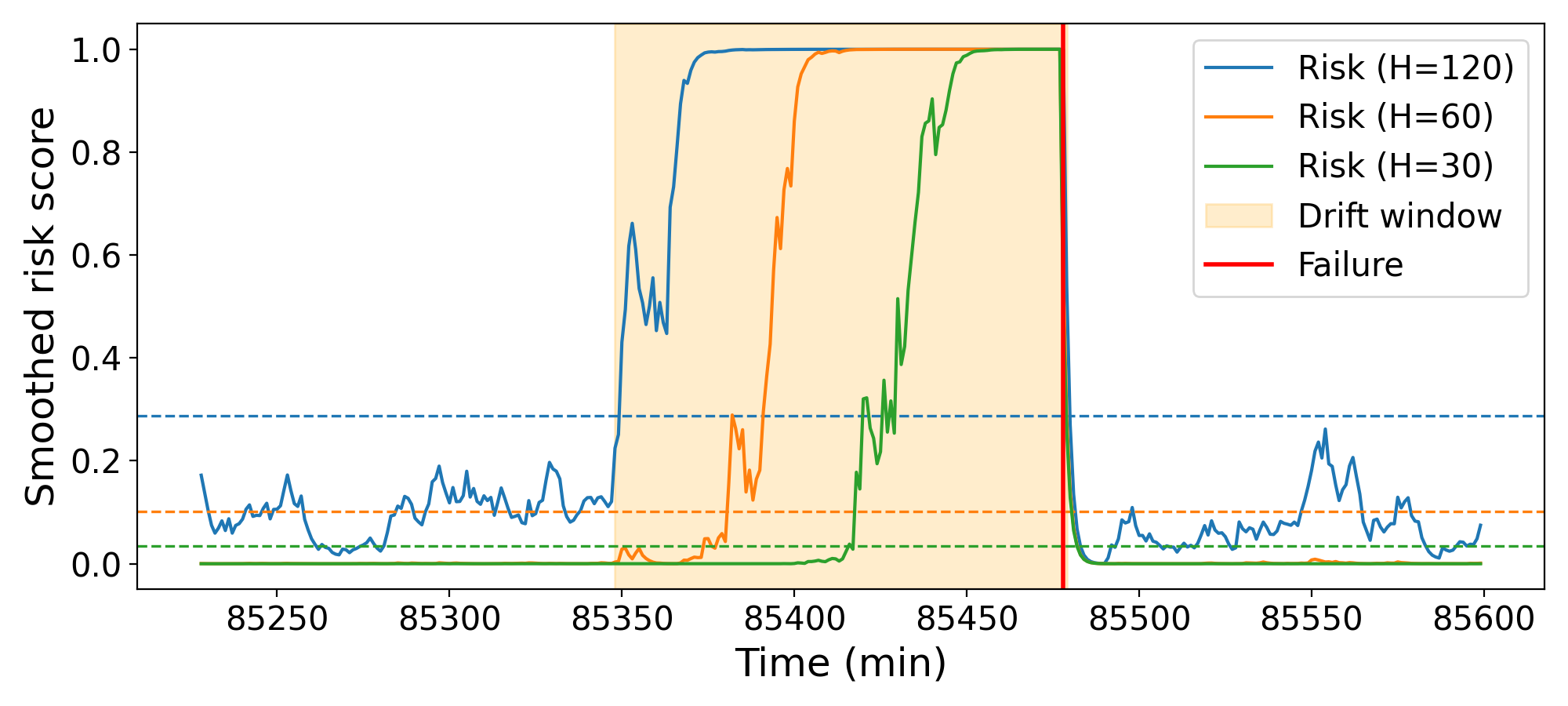}
    \caption{Multi-horizon alerting for a telemetry event (H=120,60,30).}
    \label{fig:multi_horizon}
\end{figure}

An ensemble of MILD models trained with different horizons (e.g., H=120, 60, 30 min) provides dynamic TTF estimation. In Fig.~\ref{fig:multi_horizon}, models trigger sequentially as failure approaches, tightening the window from $(60,120]$ to $(0,30]$ minutes.

\subsection{Comparison with Baselines}
We compare against five baselines trained under the same blocked CV protocol: (i) \textbf{WKPI-Tuned}, a heuristic weighted-KPI score derived from Logistic Regression coefficients; (ii) \textbf{Dist-Target}, a target-distance drift score \cite{10575429}; (iii) \textbf{LR-OvR}, per-intent OvR Logistic Regression \cite{tarekegn2024deeplearningmultilabellearning, 9592636}; (iv) \textbf{MLP}, a shared MLP encoder with per-intent sigmoid heads; (v) \textbf{LSTM}, a temporal baseline using an LSTM encoder. All methods use the same EWMA and threshold alerting policy.

\begin{figure}[t!]
    \centering
    \includegraphics[width=0.8\columnwidth]{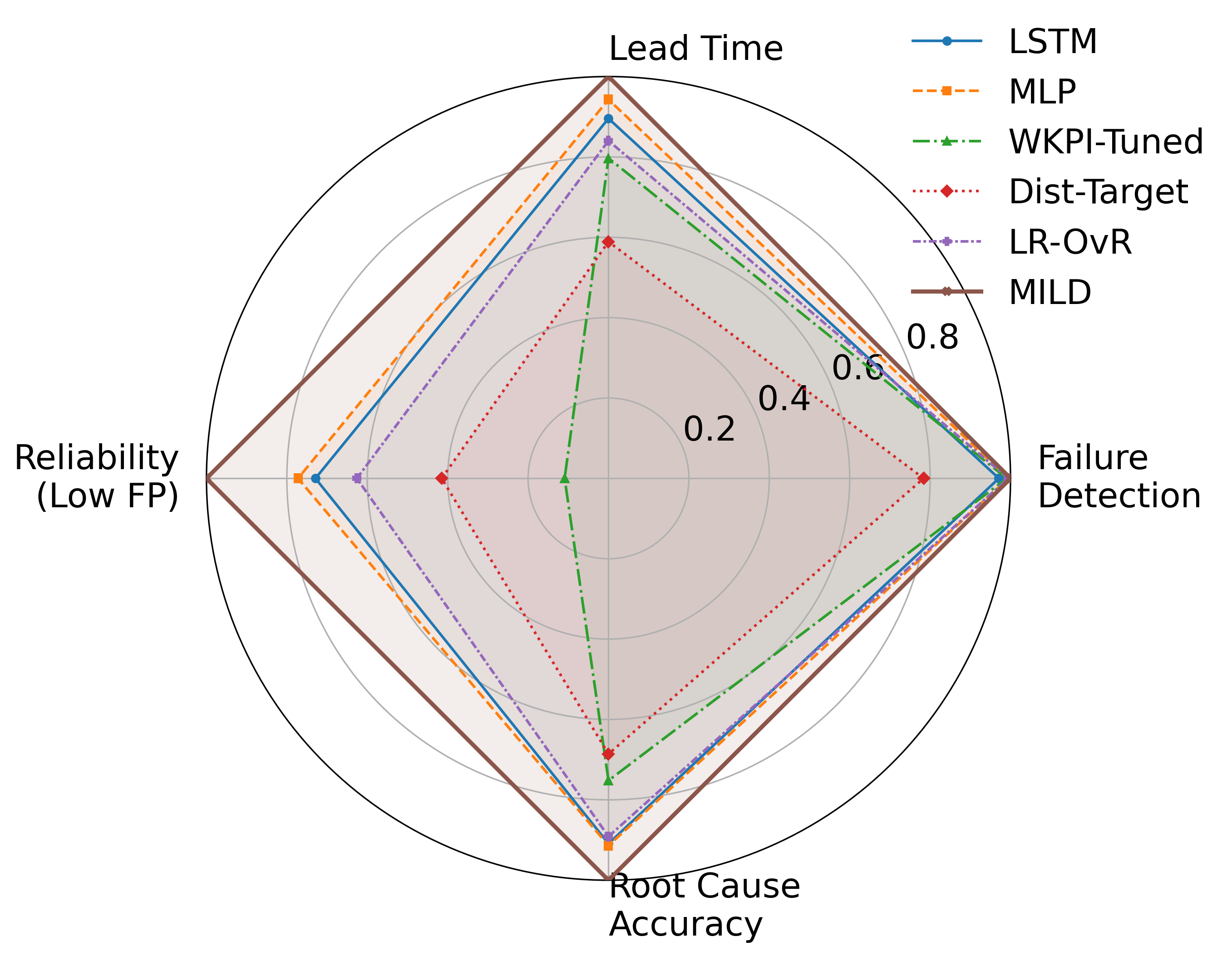}
    \caption{Normalized holistic comparison: detection, lead time, reliability (inverted FP), and root-cause accuracy.}
    \label{fig:radar}
\end{figure}

Across intents, MILD achieves the best overall trade-off with perfect detection, the longest lead times, the lowest false-positive rate, and the highest root-cause accuracy. In particular, compared to the strongest supervised baselines (LR-OvR/MLP/LSTM), MILD preserves high detection while improving proactive lead time and disambiguation. Fig.~\ref{fig:radar} summarizes the intent-averaged, percent-of-best normalized comparison across detection, lead time, reliability (inverted FP rate), and root-cause accuracy, where MILD covers the largest area.

\section{Conclusion}
We presented MILD, a framework for proactive multi-intent failure prediction with root-cause disambiguation in IBN. MILD uses a teacher-augmented Mixture-of-Experts with a trained gating network to resolve co-drift ambiguity and identify the root-cause intent. On a challenging benchmark with non-linear and co-drifting failures, MILD achieves perfect detection with long lead times, low false positives, and higher root-cause accuracy than baselines. It also provides operational intelligence via per-alert SHAP explanations and optional multi-horizon time-to-failure bounding.

\section*{Acknowledgment}
We would like to acknowledge the support provided by the Deanship of Research at King Fahd University of Petroleum \& Minerals (KFUPM).

\bibliographystyle{IEEEtran}
\bibliography{references}

\end{document}